\documentstyle[twoside,fleqn,epsfig]{article}
\def\be{\begin{equation}}
\def\ee{\end{equation}}
\def\bea{\begin{eqnarray}}
\def\eea{\end{eqnarray}}
\newcommand{\lsim}{\mathrel{\mathop{\kern 0pt \rlap
  {\raise.2ex\hbox{$<$}}}
  \lower.9ex\hbox{\kern-.190em $\sim$}}}
\newcommand{\gsim}{\mathrel{\mathop{\kern 0pt \rlap
  {\raise.2ex\hbox{$>$}}}
  \lower.9ex\hbox{\kern-.190em $\sim$}}}

\newcommand{\AmS}{{\protect\the\textfont2
  A\kern-.1667em\lower.5ex\hbox{M}\kern-.125emS}}
\normalsize
\begin{document}
\Large
\begin{flushright}
{\bf ROM2F/2007/19 \\}
{\bf to appear on Phys. Rev. D \\}
\end{flushright}
\normalsize
\vspace{-0.7cm}

\baselineskip=0.65cm
\vspace*{0.5cm}

\begin{center}
\Large \bf
Investigating electron interacting dark matter \\
\rm
\end{center}

\vspace{-0.2cm}
\vspace{0.5cm}
\normalsize

\noindent \rm R.\,Bernabei,~P.\,Belli,~F.\,Montecchia,~F.\,Nozzoli

\noindent {\it Dip. di Fisica, Universit\`a di Roma ``Tor Vergata"
and INFN, sez. Roma ``Tor Vergata", I-00133 Rome, Italy}

\vspace{3mm}

\noindent \rm F.\,Cappella, A.\,Incicchitti,~D.\,Prosperi

\noindent {\it Dip. di Fisica, Universit\`a di Roma ``La Sapienza"
and INFN, sez. Roma, I-00185 Rome, Italy}

\vspace{3mm}

\noindent \rm R.\,Cerulli

\noindent {\it Laboratori Nazionali del Gran Sasso, INFN, Assergi, Italy}

\vspace{3mm}

\noindent \rm C.J.\,Dai,~H.L.\,He,~H.H.\,Kuang, J.M.\,Ma, X.H.\,Ma,
X.D.\,Sheng, Z.P.\,Ye\footnote{also:
University of Jing Gangshan, Jiangxi, China}, R.G.\,Wang, Y.J.\,Zhang

\noindent {\it IHEP, Chinese Academy, P.O. Box 918/3, Beijing 100039, China}

\normalsize

\vspace{0.4cm}
\vspace{-0.1cm}

\begin{abstract}
\vspace{0.2cm}

Some extensions of the Standard Model provide 
Dark Matter candidate particles which can have a dominant coupling 
with the lepton sector of the ordinary matter. 
Thus, such Dark Matter candidate particles ($\chi^{0}$) can
be directly detected only through their interaction with electrons in the detectors
of a suitable experiment, while they are lost by experiments based on the rejection of the 
electromagnetic component of the experimental counting rate.  
These candidates can also offer a possible source 
of the 511 keV photons observed from the galactic bulge.
In this paper this scenario is investigated. Some theoretical arguments 
are developed and related phenomenological aspects are discussed.
Allowed intervals and regions for the characteristic phenomenological parameters 
of the considered model and of the possible mediator of the 
interaction are also 
derived considering the DAMA/NaI data.

\end{abstract}

\vspace{0.2cm}



\vspace{0.1cm} 
\vspace{-0.1cm}

\section{Introduction}

\noindent Dark Matter particles with dominant interaction on electrons have been considered
in literature \cite{Uboson,Uboson2,Uboson3,fay}. 
In particular, from a phenomenological point of view, 
Dark Matter (DM) candidates with electron interactions can offer
possible sources for the 511 keV positron annihilation line
observed from the galactic bulge \cite{integral,theor}.
These candidates can be either light (MeV scale) \cite{Uboson} or heavy 
(GeV or larger scale)
\cite{Uboson2,Uboson3}. They are expected to interact with electrons both through  
neutral light (MeV scale) U or Z' bosons
or through heavy charged 
mediators $\chi^{\pm}$ (which can eventually be nearly degenerate 
with $\chi^{0}$) \cite{Uboson3}. Recently data 
collected by some accelerator experiments
have been analyzed in terms of a $\sim 200$ MeV neutral boson which couples to quarks
with flavour changing transition: $s \rightarrow d \mu^+\mu^-$ 
\cite{B200,B200a}.
Other results showing some resonances at energies lower than the 
two-muon \cite{B200} and the two-pion \cite{B20} disintegration thresholds have 
been associated with a Goldstone neutral boson of $\sim 20$ MeV mass.
Moreover, some excess has been achieved in dedicated experiments 
on low energy nuclear reactions searching for possible $e^+-e^-$ pairs 
driven by the presence of a neutral boson with a mass around 10 MeV
\cite{B10}.

Let us remark that -- in the frameworks where the mediator is either a $\pm 1$ charged boson or a neutral boson 
providing a flavour changing transition among quarks -- the elastic scatterings of 
the DM candidate $\chi^0$ particles on nuclei 
would be either forbidden or suppressed; hence, the scattering on electrons 
would remain the unique possibility for the direct detection of the $\chi^0$ particles.

On the other hand, from a pure theoretical point of view, it is also conceivable 
that the mediator of the DM particle interactions can be coupled only 
to the lepton sector of the ordinary matter. Thus, in this case the DM particles can 
just interact with electrons and cannot with nuclei. This is suggested in 
ref. \cite{fay} for the U boson and can also be the case of 
some extensions \footnote{For example
from the extended Pati-Salam gauge group $SU(6) \times SU(2)_L \times 
SU(2)_R$ \cite{patisalam} or from $[SU(3)]^4$ quartification 
\cite{quartification}.} 
of the Standard Model providing a quark-lepton discrete symmetry
$SU(3)_l \times SU(3)_q \times SU(2)_L \times U(1)$.
In these latter models, leptons (as well as quarks)
are assumed to have three "leptonic ($l$) colours" and to interact
through the gauge group $SU(3)_l$, analogously as the QCD colour group 
$SU(3)_q$. Moreover, at some high energy scale 
a symmetry breaking $SU(3)_l \rightarrow SU(2)'$ is expected, 
giving high mass to the "exotic" leptonic degree of freedom and 
leaving light the "standard" leptons \cite{lcolor}.
In these scenarios, the heavy "exotic" leptonic degree of freedom 
provides both heavy charged $\pm 1/2$ fermions, which are expected to 
be confined into exotic {\it leptonic hadrons} by the unbroken gauge group $SU(2)'$ \cite{lcolor},
and heavy neutrinos \cite{quartification,lcolor}; hence, they 
can be considered as Dark Matter candidates with dominant interaction on electrons.

Moreover, it is worth to note that other possibilities can exist. For example,  
supersymmetric (SUSY) theories can offer configurations in the general 
SUSY parameter space where the lightest supersymmetric particle (LSP) 
has an interaction with electron dominant with respect to that with quark.

These DM candidate particles can be directly detected only through their interaction with electrons 
in the detectors of a suitable experiment, while they are lost by experiments based on the rejection of the 
electromagnetic component of the experimental counting rate.  

In the present paper this kind of DM candidates are investigated,
some theoretical arguments are developed and
related phenomenological aspects are discussed.
In particular, the impact of these DM candidates will also be discussed in a
phenomenological framework on the basis of the 6.3 $\sigma$ C.L. DAMA/NaI 
model independent evidence for particle Dark Matter in the galactic 
halo \cite{RNC,ijmd}. We remind that various corollary analyses, considering  
some of the many possible astrophysical, nuclear and particle Physics scenarios,
have been analysed by DAMA itself both for some WIMP/WIMP-like candidates and for light bosons 
\cite{RNC,ijmd,epj06,ijma2,ijma,chann}, 
while several others are also available in literature, such as e.g.
refs. \cite{Bo03,Bo04,Botdm,khlopov,Wei01,foot,Saib,droby1,droby2,sneu,zoom}. 
Many other scenarios can be considered as well.
At present the new second generation DAMA/LIBRA set-up is running at the Gran Sasso
Laboratory.

\vspace{0.5cm}

\section{Detectable energy in $\chi^{0}$ - electron elastic scattering}

\begin{figure} [!ht]
\centering
\vspace{-0.2cm}
\includegraphics[width=180pt] {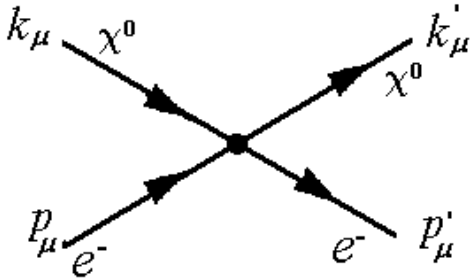}
\vspace{-0.2cm}
\caption{The $\chi^0$ -- $e^-$ elastic scattering and definition
of the momentum variables in the laboratory frame.
In the text a contact interaction has been assumed (also see Appendix B) as 
suitable approximation of the process.}
\label{fg:diagram_we}
\end{figure}

The practical possibility to detect electron interacting DM candidates
(hereafter $\chi^{0}$ with mass $m_{\chi^0}$ and 4-momentum $k_\mu$)
is based on the detectability of the energy released in 
$\chi^{0}$ - electron elastic scattering processes (see Fig. \ref{fg:diagram_we}). 

Generally, these processes are not taken into account in the DM field since the electron is
assumed at rest and, therefore, considering the $\chi^0$ particle 
velocity $|\vec{v}_{\chi^0}| \sim 300$ km/s, the released energy is of the order of few eV, well below
the detectable energy in any considered detector in the field. However,
the electron is bound in the atom and, even
if the atom is at rest, the electron can have not negligible momentum, $p$.
For example, the bound electrons in NaI(Tl) offer a probability 
equal to $\sim 1.5 \times 10^{-4}$ to have $p \gsim$ 0.5 MeV/c;
such a probability is quite small, but not zero.
Hence, interactions of $\chi^0$ particles with these high-momentum electrons 
in an atom at rest can give rise to detectable signals in suitable detectors. 
In particular, after the interaction the final state can have -- beyond the scattered $\chi^0$ particle --
either a prompt electron and an ionized atom or an excited atom plus possible X-rays/Auger electrons.
Therefore, the process produces X-rays and electrons of relatively low energy,
which are mostly contained with efficiency $\sim 1$ in a detector of a suitable size.
Thus, the total detected energy, 
$E_d = k_0 - k_0' = p_0' - p_0$ (where $k_0$, $k_0'$, $p_0'$ and $p_0$ are the time components of the
respective 4-vectors in the laboratory frame, 
see Fig. \ref{fg:diagram_we}), can be evaluated considering 
the energy conservation in the centre of mass (CM) frame of the $\chi^0-e^-$ system. 
Defining $\vec{\beta} = \frac{\vec{k}+\vec{p}}{k_0+p_0}$
as the velocity of the 
CM frame with the respect to the laboratory frame and $\gamma=1/\sqrt{1-\beta^2}$ Lorentz
boost factor, one can write the energies
of the electron before and after the scattering
by using the variables in the CM frame through the Lorentz transformations:
\begin{equation}
p_0=\gamma (p_{0,CM} + \vec{\beta} \cdot \vec{p}_{CM}) 
\,\,\,\,\,\,\,  \mbox{and}  \,\,\,\,\,\,\, 
p'_0=\gamma (p'_{0,CM} + \vec{\beta} \cdot \vec{p'}_{CM}) .
\end{equation}
Since we are dealing with elastic scattering, $p_{0,CM}=p'_{0,CM}$ and 
$|\vec{p}_{CM}|=|\vec{p'}_{CM}|$, so that, by subtraction, one obtains:
\begin{eqnarray}
E_d = \gamma \left( \vec{\beta} \cdot \vec{p'}_{CM}-
\vec{\beta} \cdot \vec{p}_{CM} \right) = \gamma \; \beta \; p_{CM} 
 \left( cos \theta' - cos \theta \right)
\label{eq:ED}
\end{eqnarray}
where $\theta'$ is the angle between $\vec{\beta}$ and $\vec{p'}_{CM}$,
$\theta$ is the angle between $\vec{\beta}$ and $\vec{p}_{CM}$ and
$\vec{p}_{CM} = \gamma (\vec{p} - \vec{\beta} p_0)$.

\begin{figure} [!ht]
\centering
\vspace{-0.2cm}
\includegraphics[width=180pt] {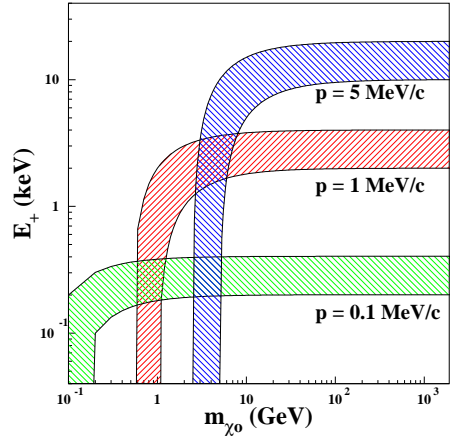}
\vspace{-0.5cm}
\caption{Few examples of the dependence of the maximum released energy, $E_+$, 
on the $\chi^0$ mass for electron's momenta of
0.1, 1 and 5 MeV/c, for $v_{\chi^0}$ ranging in the interval $1 \div 2 \times 10^{-3}c$
and for head-on collisions ($\theta=\pi$).}
\label{fg:emax}
\end{figure}

Therefore, fixing the input momenta of the $\chi^0$ particle ($\vec{k}$) and 
of the electron ($\vec{p}$), the maximum detected energy is given by:
$E_+ = \gamma \; \beta \; p_{CM} (1 - cos \theta )$. Few examples of 
the dependence of $E_+$ on the $\chi^0$ mass are given in Fig. \ref{fg:emax}
as function of the electron's momentum and of the $\chi^0$ velocities 
for head-on collisions ($\theta=\pi$).
The  Fig. \ref{fg:emax} also points out that $\chi^0$ particles with 
$m_{\chi^0}$ larger than few GeV can provide 
sufficient energy to be detected in a suitable detector.

It is interesting to explore two limit cases (remind that owing to the typical 
$\chi^0$ velocities, $k_0 \simeq m_{\chi^0}$ and $\vec{k} \simeq m_{\chi^0} \cdot \vec{v}_{\chi^0}$;
hereafter $c=1$):

\newcounter{a}
\begin{list}{\alph{a})}{\usecounter{a}}

\item $p \ll \beta m_e \sim$ keV, that is target 
nearly at rest\footnote{We note that in general for a target of mass $m_T$ nearly at rest: 
$E_+ \simeq 2 \beta^2 m_T = $
\mbox{$\frac{1}{2} m_{\chi^0} v^2_{\chi^0} \cdot \frac{4 m_{\chi^0} m_T}{(m_{\chi^0} + m_T)^2}$},
that is one gets the formula describing for example the WIMP-nucleus elastic scattering.}: 
$E_+ \simeq 2 \beta^2 m_e \sim$ eV.

\item $k \gg p \gg \beta p_0 \sim$ keV; in this case
one obtains $\vec{p}_{CM} \simeq \vec{p}$, $\vec{\beta} \simeq \vec{v}_{\chi^0}$
and, therefore, $\theta$ is also the angle between $\vec{p}$ and $\vec{k}$. Hence:
$E_+ \simeq v_{\chi^0} p (1 - cos \theta )$. This is the case
of interest for the direct detection; in fact, for $m_{\chi^0}$ larger than few GeV 
$k$ is larger than the maximum momentum of a bound electron in the atom 
due to the finite size of the nucleus ($\sim 15$ MeV in Iodine). 

\end{list}

In conclusion, $\chi^0$ particles with mass 
$\gsim$ few GeV, interacting on bound electrons with 
momentum up to $\simeq$ few MeV/c (see case b), can provide signals in the keV 
energy region detectable by low background 
and low energy threshold detectors, such as those of DAMA/NaI (see later).

\section{Cross section and counting rate}

\subsection{The cross section at fixed electron momentum}

The differential cross section for $\chi^{0}$ - electron elastic scattering can be written as:

\begin{equation}
d\sigma = \frac{\overline{|M|^2}}{v_{(\chi^0 e)}}
\frac{1}{2k_02p_0}
(2\pi)^4\delta^4(k+p-k'-p')
\frac{d^3p'}{(2\pi)^32p'_0}
\frac{d^3k'}{(2\pi)^32k'_0} \, .
\label{eq:d1}
\end{equation}

\noindent There $\overline{|M|^2}$ is the averaged squared matrix element and
$v_{(\chi^0 e)}$ is the relative velocity between $\chi^{0}$ and the electron.

\noindent Integrating over $d^3k'$ and over the $p'$ solid angle and 
considering that $p' dp' = p'_0 dp'_0 = p'_0 dE_d$, one can write:
\begin{equation}
\frac{d\sigma}{dE_d} =
\frac{\overline{|M|^2}}
{32\pi v_{(\chi^0 e)} k_0 p_0} \cdot
\frac{1}{|\vec{k}+\vec{p}|} \cdot \theta(E_+ - E_d) \, .
\label{eq:sv33}
\end{equation}
The Heaviside theta function defines the domain of the differential cross section.

It is useful in the following to define the $\chi^0$ cross section on the electron 
at rest ($p = 0$); thus, one can write:

\begin{equation}
\left. \frac{d\sigma}{dE_d} \right|_{(p=0)}
= 
\frac{\overline{|M|^2}_{(p=0)}}
{32 \pi v_{\chi^0} k_0 m_e}
\frac{1}{k}
\theta(E_{+}-E_d)=\frac{\sigma_e^0}{E_{+}}\theta(E_{+}-E_d) ,
\label{eq:sv55}
\end{equation}
where $E_{+} (p=0) = 2m_e v_{\chi^0}^2 \sim eV$
and $\sigma_e^0 = \frac{\overline{|M|^2}_{(p=0)}}
{16\pi m_{\chi^0}^2}$.
In the following, for simplicity, we define $\sigma_e = 
\frac{\overline{|M|^2}} {16\pi m_{\chi^0}^2}$, then 
$\sigma_e(p=0)=\sigma_e^0$.

\subsection{The cross section for atomic electrons}

Let us now introduce in the previous evaluations the momentum 
distribution of the electrons in the atom, $\rho(\vec{p})$ (see Appendix A).
In particular, from eq. (\ref{eq:sv33}) -- that is for a fixed $\vec{p}$ value --
one can write for the atomic case:
\begin{equation}
\frac{d\sigma}{dE_d} = 
\frac{\overline{|M|^2}}
{32 \pi v_{(\chi^0 e)} k_0 p_0}
\frac{1}{|\vec{k}+\vec{p}|}\theta(E_+-E_d) \rho(\vec{p}) d^3p \,.
\label{eq:sv44}
\end{equation}
Introducing the $\sigma_e$ definition and replacing
$E_{+}$ with its expression, it is possible to write for 
the relevant case of direct detection ($k \gg p \gg m_e v_{\chi^0} $):
\begin{equation}
\frac{d\sigma}{dE_d} \simeq 
\frac{\sigma_e p^2}
{2 v_{(\chi^0 e)} v_{\chi^0} p_0}
\rho(\vec{p}) d\phi dcos\theta
\;
\theta[v_{\chi^0} p(1-cos\theta)-E_d] dp ;
\label{eq:sv4}
\end{equation}
here the polar axis has been chosen in the direction of 
$\vec{k}$.

The integration over $\phi$ simply gives $2\pi$ considering that 
$\overline{|M|^2}$ does not depend on $\phi$ and that atoms with full 
shells (as $Na^+$ and $I^-$) have isotropic distributions $\rho(p)$.

\subsection{The counting rate}

The expected interaction rate of $\chi^0$ particle impinging on the electrons
of an atom can be derived as:

\begin{equation}
\frac{dR}{dE_d} = 
\frac{\rho_{\chi^0}}{m_{\chi^0}} \eta_e
\int \frac{d\sigma}{dE_d} v_{(\chi^0 e)}
f(\vec{v}_{\chi^0})d^3v_{\chi^0} ,
\label{eq:rate}
\end{equation}

\noindent where: i) $\rho_{\chi^0} = \xi \rho_0$ with $\rho_0$ 
local halo density and $\xi \le 1$ fractional amount 
of $\chi^0$ density in the halo; ii) $f(\vec{v}_{\chi^0})$ is the $\chi^0$
velocity ($v_{\chi^0}$) distribution in the Earth frame;
iii) $\eta_e$ is the electron's number density in the target material.  

In the reasonable hypothesis that $\sigma_e$ does not depend on $cos\theta$,
the integrand in eq. (\ref{eq:rate}) can be evaluated
considering that:
\begin{equation}
\frac{d\sigma}{dE_d} \cdot v_{(\chi^0 e)} =
\frac{2\pi \sigma_e p^2}
{v_{\chi^0}^2 p_0}
\rho(p) (v_{\chi^0} - v_{min})
\theta (v_{\chi^0} - v_{min}) dp \; ,
\label{eq:rate2}
\end{equation}
where $v_{min}=\frac{E_d}{2p}$ is the minimal $\chi^0$ particle velocity 
in order to provide an energy $E_d$ released in the detector.

The matrix element $|M|^2$ -- as well as $\sigma_e$ in eq. (\ref{eq:rate2}) --  
can generally depend on $p$ and $v_{\chi^0}$.
Thus, in order to evaluate it, it is necessary to consider 
a specific particle interaction model (see Appendix B).

For simplicity, we will consider a 4-fermion contact interaction 
(e.g. a mediator with mass larger than many MeV, neglecting the 
4-momentum transferred into the propagator). Thus,
for the cases of pure $V \pm A$ and pure scalar interactions --
which are addressed in the following -- one gets: $\sigma_e \simeq \sigma_e^0 
\frac{p_0^2}{m_e^2}$.
Other interaction models are possible and can be investigated 
in the future. It is worthwhile to stress that 
-- although the calculations are made for the $V\pm A$ and 
for the scalar 4-fermion contact interactions -- 
same results can be achieved for any kind of DM candidate interacting with electrons
and with cross section $\sigma_e$ having a weak dependence on $p$ and $v_{\chi^0}$,
that is $\sigma_e \sim \sigma_e^0$.

Finally, the expected interaction rate can be written as:
\begin{equation}
\frac{dR}{dE_d} = 
\frac{\xi \sigma_e^0}{m_{\chi^0}} \cdot
\frac{2\pi \rho_{0}}{m_e^2} \eta_e 
\int_0^\infty p^2 p_0 \rho(p) \cdot
I(v_{min}) \; dp \, ,
\label{eq:rate33}
\end{equation}
where -- pointing out the time dependence of $f(\vec{v}_{\chi^0})$ --
we have introduced the useful function:
\begin{equation}
I(v_{min})= \int_{v_{min}}^{\infty}
\frac{f(\vec{v}_{\chi^0})}{v_{\chi^0}^2}
(v_{\chi^0} - v_{min})
d^3v_{\chi^0} 
\simeq I_{0}(v_{min}) + I_{m}(v_{min}) \cdot cos \omega (t-t_0).
\label{eq:ivmin}
\end{equation}
Here roughly $t_0 \simeq 2^{nd}$ June and $\omega = \frac{2\pi}{T}$ with $T = 1$ yr.
The cut-off of the halo escaping velocity is included into 
the $f(\vec{v}_{\chi^0})$ function distribution.

Therefore, the expected counting rate accounting for the energy resolution
of the detector can be written as:

\begin{equation}
\frac{dR}{dE} = \int G(E,E_d)\frac{dR}{dE_d}dE_d = S_0 + S_m \cdot cos \omega (t-t_0) \, ,
\label{eq:rate333}
\end{equation}
where $S_0$ and $S_m$ are the unmodulated and the modulated part of the expected signal, 
respectively. The $G(E,E_d)$ kernel generally has a gaussian behaviour.

Finally, we
note that -- since $m_{\chi^0}$ is larger than few GeV (so that $k \gg p$) -- 
the expected counting rate has a simple dependence upon $\sigma_e^0$ and $m_{\chi^0}$;
therefore, the ratio $\frac{\xi \sigma_e^0}{m_{\chi^0}}$ is a normalization factor
of the expected energy distribution.

\begin{figure} [!h]
\centering
\vspace{-0.4cm}
\includegraphics[width=160pt] {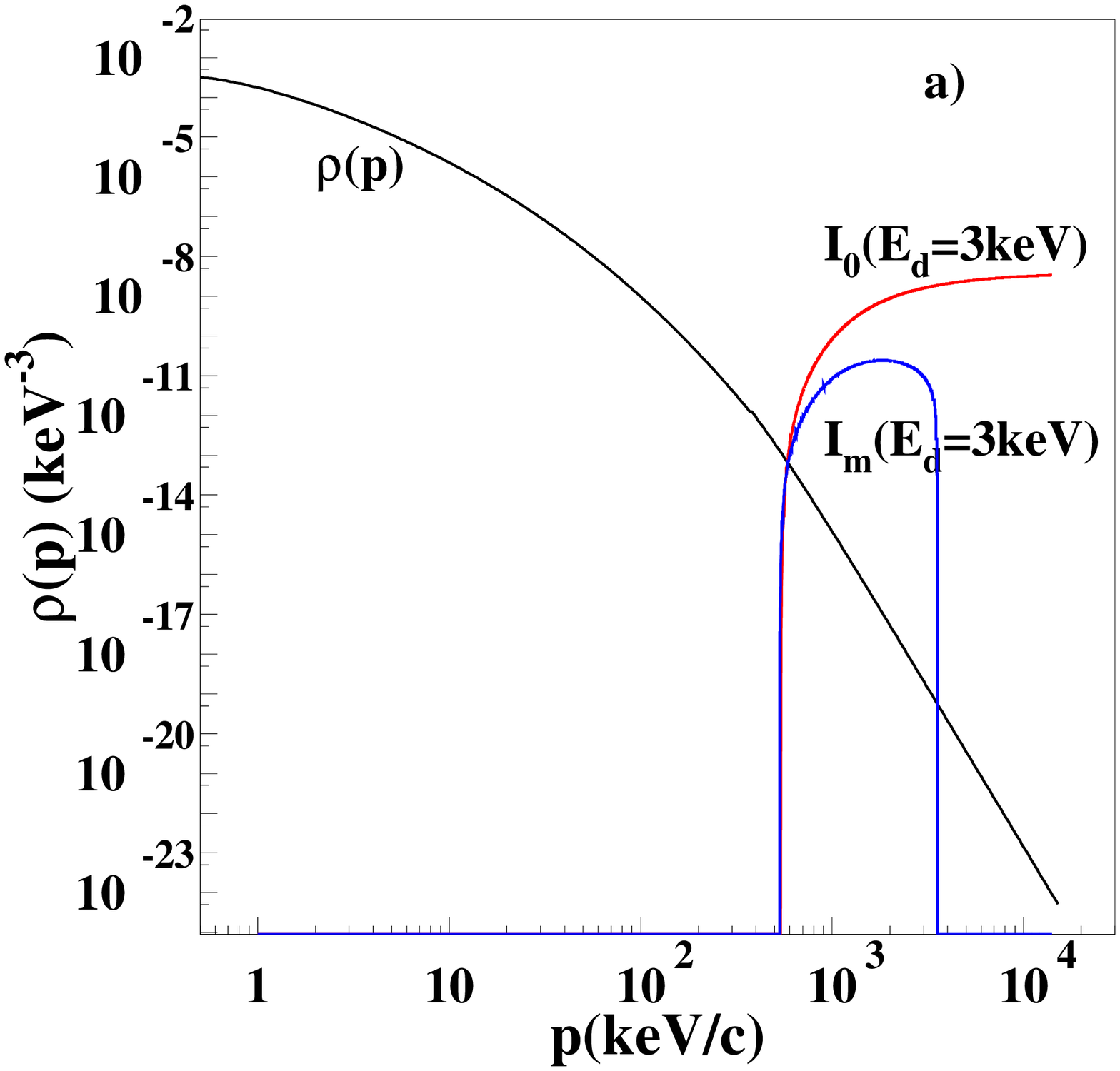}
\includegraphics[width=160pt] {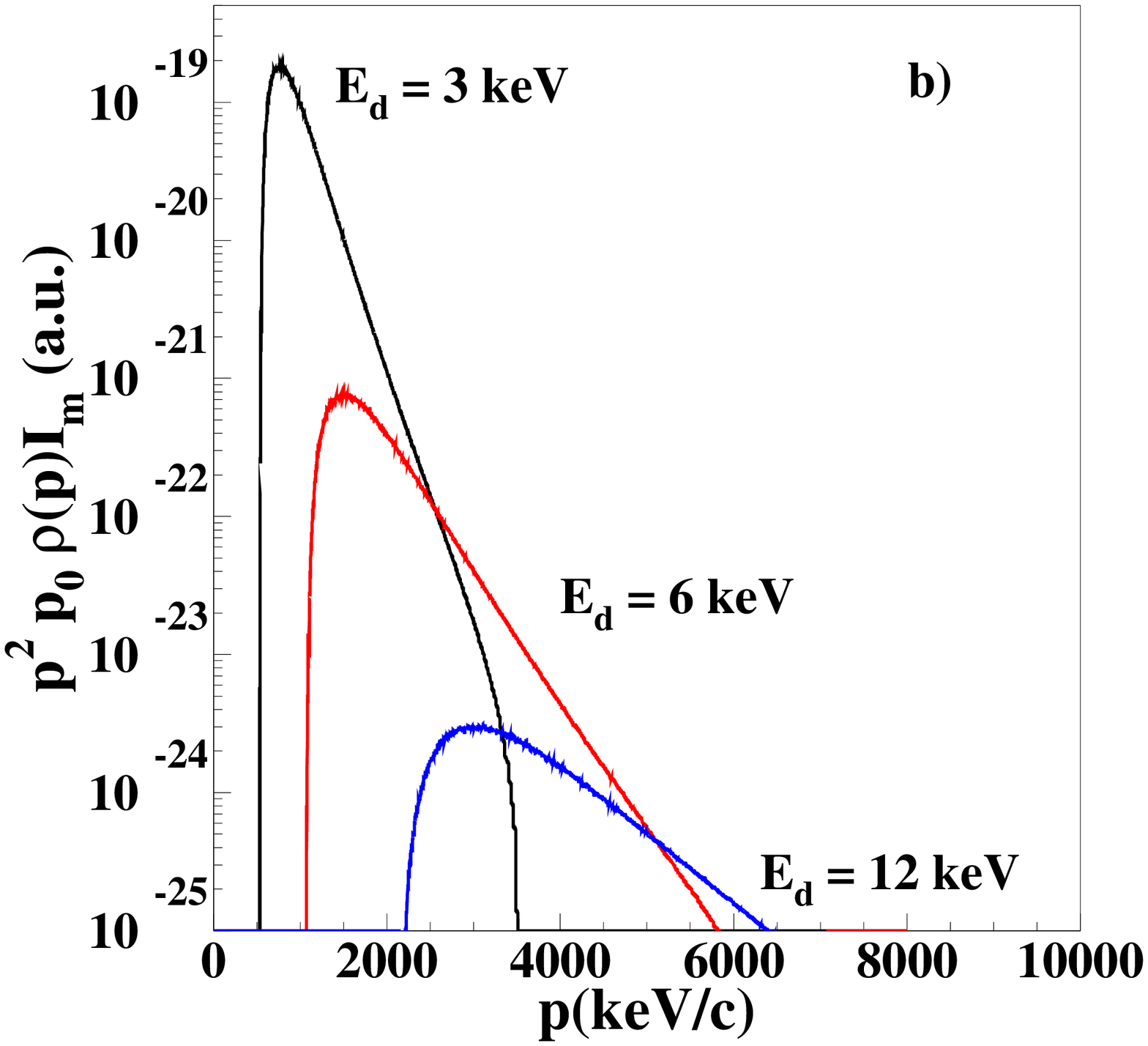}
\vspace{-0.4cm}
\caption{a) Behaviours of $\rho(p)$ (solid black line) for NaI(Tl)
and $I_0$ and $I_m$ for $E_d = 3$ keV in the considered halo model, 
A5 of ref. \cite{Hep,RNC}; see also text. The functions $I_0$ and $I_m$ are in arbitrary units.
b) Behaviours of $p^2 p_0 \rho(p) I_m$ for NaI(Tl) at three different values of the
released energy: $E_d = $ 3, 6 and 12 keV 
in the considered halo model, A5 of ref. \cite{Hep,RNC};
they show as the main contribution to the counting rate in NaI(Tl) detectors with 
energy threshold at 2 keV comes from electrons with momenta around few MeV/c.}
\label{fg:fig5q}
\end{figure}

The momentum distribution of the electrons in NaI(Tl), $\rho(p)$,
has been depicted in Fig. \ref{fg:fig5q}a); it has been
calculated from the corresponding Compton profile, $J(p)$,
reported in ref. \cite{biggs}. For this purpose, due to the isotropic 
distributions of Na$^+$ and I$^-$ (ions with full shells) 
the relation $J(p) = 2\pi \int_p^{\infty} \rho(q) q dq$ has been used  
\cite{cprof2,cprof3}. At high momentum the $\rho(p)$ function follows
the hydrogenic behaviour of the $1s$ internal shell of the Iodine atom:
$\rho(p) \propto \left(p_I^2+p^2 \right)^{-4}$ 
with $p_I \simeq 200$ keV. 

As an example, in Fig. \ref{fg:fig5q}a) the behaviours of $I_0(v_{min})$, $I_m(v_{min})$ 
and $\rho(p)$ are compared as function of the electron's momentum,
$p$, for NaI(Tl) as target material and for the given released energy: $E_d=3$ keV. 
In this figure as template the considered halo model is the A5 model of ref. \cite{Hep,RNC}, that is 
a NFW halo model with local velocity equal to 220 km/s and density equal to the maximum value 
($\rho_0$ = 0.74 GeV cm$^{-3}$).

\begin{figure} [!h]
\centering
\vspace{-0.8cm}
\includegraphics[width=160pt] {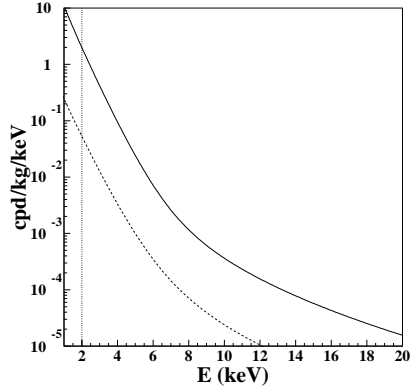}
\vspace{-0.4cm}
\caption{An example of the shapes of expected energy distributions in NaI(Tl)
due to $\chi^0$ interactions with electrons for the scenario given in the text;
the solid line gives the behaviour of the unmodulated part of the expected signal, $S_0$,
while the dashed line is the behaviour of the modulated part, $S_m$.
In this example the normalization factor 
is $\frac{\xi \sigma^0_e}{m_{\chi^0}} = 7 \times 10^{-3}$ pb/GeV.
The vertical line indicates the energy threshold of the DAMA/NaI experiment.}
\label{fg:fig6q}
\vspace{-0.2cm}
\end{figure}

It is possible to see that -- due to the behaviour of the momentum distribution 
of the electrons, $\rho(p)$, at high $p$ and due to the behaviour of the $I$ 
function at low $p$ (related to the $f(\vec{v}_{\chi^0})$ behaviour at high velocity) --
the main contribution to the counting rate in NaI(Tl) detectors with 
energy threshold at 2 keV
comes from electrons with momenta around few MeV/c
(see Fig. \ref{fg:fig5q}b). It is worthwhile to note that similar behaviours can also be obtained
by using other choices of the halo model.

Finally, an example of the shapes of expected energy distributions in NaI(Tl)
due to $\chi^0$ interactions with electrons for the A5 halo model (a NFW halo 
model with local velocity equal to 220 km/s and density equal to the maximum value,
see ref. \cite{Hep,RNC}) is reported in Fig. \ref{fg:fig6q}.
In this example 
the normalization factor is $\frac{\xi \sigma^0_e}{m_{\chi^0}} = 7 \times 10^{-3}$ pb/GeV.

\section{Data analysis and results for electron interacting DM candidate in DAMA/NaI}

The 6.3 $\sigma$ C.L. model independent  
evidence for Dark Matter particles in the galactic halo 
achieved over seven annual cycles by DAMA/NaI \cite{RNC,ijmd} 
(total exposure $\simeq 1.1 \times 10^{5}$ kg $\times$ days) 
can also be investigated for the case of an electron interacting DM 
candidate (in addition 
to the other corollary quests already mentioned in the previous 
footnote 4). 

In the analysis presented here, the same dark halo models 
and related parameters given in table VI of ref. \cite{RNC} have been used;
the related DM density is given in table VII of the same reference. Moreover, here
$\eta_e$ = 2.6 $\times$ 10$^{26}$ kg$^{-1}$ and the halo escaping velocity has been taken 
equal to 650 km/s.

The results are calculated by taking into
account the time and energy behaviours of the {\it single-hit} experimental data
through the standard maximum likelihood method\footnote{Shortly, the likelihood function is: 
${\it\bf L}  = {\bf \Pi}_{ijk} e^{-\mu_{ijk}}
{\mu_{ijk}^{N_{ijk}} \over N_{ijk}!}$, where
$N_{ijk}$ is the number of events collected in the
$i$-th time interval, by the $j$-th detector and in the
$k$-th energy bin. $N_{ijk}$ follows a Poissonian
distribution with expectation value 
$\mu_{ijk} = [b_{jk} + S_{0,k} + S_{m,k} \cdot  cos\omega(t_i-t_0)] M_j \Delta
t_i \Delta E \epsilon_{jk}$. 
The unmodulated and modulated parts of the signal,
$S_{0,k}$ and $S_{m,k}cos\omega(t_i-t_0)$, respectively,
are here functions of the only free parameter of the fit:
the $\frac{\xi\sigma_e^0}{m_{\chi^0}}$ ratio.
The b$_{jk}$ is the background contribution;
$\Delta t_i$ is the detector running time during the $i$-th time interval; 
$\epsilon_{jk}$ is the overall efficiency and $M_j$ is the detector mass.}.
In particular, they are presented in terms of the allowed interval
of the $\frac{\xi\sigma_e^0}{m_{\chi^0}}$ parameter, 
obtained as superposition of the configurations corresponding
to likelihood function
values {\it distant} more than $4\sigma$ from
the null hypothesis (absence of modulation) in each one of the several
(but still a very limited number) of the considered model frameworks.
This allows us to account for at least some of the 
existing theoretical and experimental uncertainties  
(see e.g. in  ref. \cite{RNC,ijmd,epj06,ijma2,ijma,chann} and in 
literature). 

\begin{figure} [!ht]
\centering
\vspace{-1.cm}
\includegraphics[width=180pt] {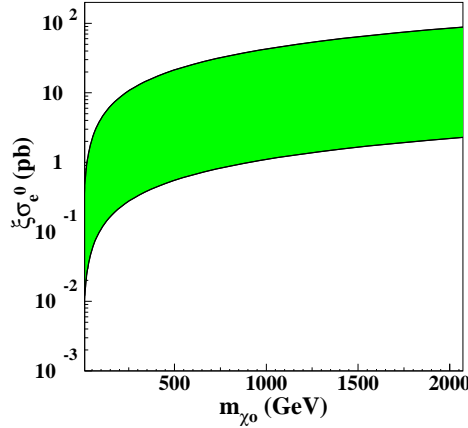}
\vspace{-0.6cm}
\caption{The DAMA/NaI region allowed in the ($\xi \sigma_e^0$ vs $m_{\chi^0}$) plane
for the same dark halo models and  
related parameters described in ref. \cite{RNC}. 
The region encloses configurations corresponding to likelihood function values
{\it distant} more than $4\sigma$ from
the null hypothesis (absence of modulation).
We note that, although the mass region in the plot is up to 2 TeV,
$\chi^0$ particles with larger masses are also allowed.}
\label{fg:reg}
\end{figure}

For these scenarios the DAMA/NaI annual modulation data gives for the considered $\chi^0$ 
candidate:
$1.1 \times 10^{-3}$ pb/GeV $< \frac{\xi \sigma_e^0}{m_{\chi^0}}
< 42.7 \times 10^{-3}$ pb/GeV at $4\sigma$ from null hypothesis.
In particular, Fig. \ref{fg:reg} shows the DAMA/NaI region allowed in the
($\xi \sigma_e^0$ vs $m_{\chi^0}$) plane for the same 
dark halo models and related parameters described in ref. \cite{RNC}.

We would like to stress that -- although the above mentioned calculations 
have been made for the $V\pm A$ and for the scalar 4-fermion contact interactions -- 
the results given here hold for every kind of DM candidate interacting with electrons
and with cross section $\sigma_e$ having a weak dependence on $p$ and 
$v_{\chi^0}$,
that is $\sigma_e \sim \sigma_e^0$; in such a case, the DAMA/NaI annual modulation data
gives: $1.6 \times 10^{-3}$ pb/GeV $< \frac{\xi \sigma_e^0}{m_{\chi^0}}
< 53.4 \times 10^{-3}$ pb/GeV at $4\sigma$ from null 
hypothesis.

Let us now comment some phenomenological implications about the possible
mediator of the interaction (hereafter U boson). The hypothesis of
4-fermion contact interaction still holds for U boson masses, 
$M_U$, larger than the transferred momentum ($M_U \gsim 10$ MeV).
In the pure $V\pm A$ and pure scalar scenario, the 
cross section is given by (see Appendix B):
\begin{equation}
\sigma_e^0 = \frac{\overline{|M|^2}} {16\pi m_{\chi^0}^2}
= \frac{16 G^2 m_{\chi^0}^2 m_e^2} {16\pi m_{\chi^0}^2}
= \frac{G^2m_e^2}{\pi} = \frac{c_e^2 c_{\chi^0}^2 m_e^2}{\pi M_U^4} .
\label{eq:propa}
\end{equation}
The effective coupling constant, $G$, depends on the couplings, $c_e$ and $c_{\chi^0}$,
of the U boson with the electron and the $\chi^0$ particle, respectively. 
We note that limits on $c_e$ have been achieved by the experimental constraints 
on the possible U boson coupling to electron arising from the $g_e-2$ measurements: 
$c_e \lsim 10^{-4} \frac{M_U}{MeV}$ \cite{fay}. 
Moreover, more restrictive limits have been obtained under the assumption 
of universality ($c_{\mu} \sim c_e \sim c_{\nu}$) by
considering the $g_{\mu}-2$ and $\nu - e$ scattering data:
$ \lsim 3 \times 10^{-6} \frac{M_U}{MeV}$ \cite{fay}.
 
\vspace{0.3cm}

The DAMA/NaI allowed region of Fig. \ref{fg:reg} requires values of $c_e$
well in agreement with these experimental upper limits. In fact,
from Fig. \ref{fg:reg} and reminding that $\xi \le 1$ and $m_{\chi^0} \gsim$ few GeV
(see above), we obtain that $\sigma_e^0 \gsim 10^{-2}$ pb.
Requiring that the theory remains perturbative (that is, $c_{\chi^0} < \sqrt{4\pi}$)
and for $M_U \sim 10$ MeV, the values of $c_e$ allowed by DAMA/NaI data
are (see eq. (\ref{eq:propa})): 
$c_e \gsim 5 \times 10^{-7}$, in agreement with the experimental upper limits.

\begin{figure} [!hb]
\centering
\vspace{-0.5cm}
\includegraphics[width=170pt] {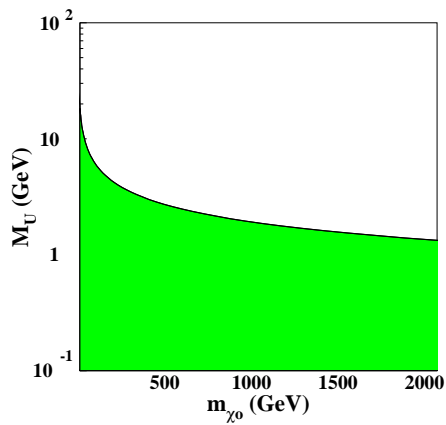}
\vspace{-0.4cm}
\caption{Region of U boson mass allowed by present 
analysis and by the $g_e-2$ constrain \cite{fay} considering that
$\xi \le 1$ and that the theory is
perturbative ($c_{\chi^0} < \sqrt{4\pi}$). See text.
There U boson with $M_U$ masses in the sub-GeV range 
required by the analyses of ref. \cite{Uboson,fay,B200,B200a} is well allowed 
for a large interval of $m_{\chi^0}$.}
\label{fg:uboson}
\end{figure}

More in general, considering the limit on $c_e$ from $g_e-2$ data and 
the obtained lower bound $\frac{\xi \sigma_e^0}{m_{\chi^0}} > 1.1 \times 10^{-3}$ pb/GeV
from the DAMA/NaI data, the allowed U boson masses are:
$M_U (GeV)  \lsim \sqrt{\frac{3700}{m_{\chi^0}(GeV)}}$, as reported in 
Fig. \ref{fg:uboson}. 
There U boson with $M_U$ masses in the sub-GeV range 
required by the analyses of ref. \cite{Uboson,fay,B200,B200a,B20,B10} is well allowed 
for a large interval of $m_{\chi^0}$.

\section{Conclusions}

In this paper, the scenario of a DM particle $\chi^{0}$ with
dominant interaction with electrons has been investigated. 
This candidate can be directly detected only
through its interaction with electrons in suitable detectors. 
Theoretical arguments have been developed and
related phenomenological aspects have been discussed.
In particular, the impact of these DM candidates has also been analysed 
in a phenomenological framework on the basis of the DAMA/NaI data.

For the considered dark halo models the
DAMA/NaI data support for the $\chi^0$ candidate:
$1.1 \times 10^{-3}$ pb/GeV $< \frac{\xi \sigma_e^0}{m_{\chi^0}}
< 42.7 \times 10^{-3}$ pb/GeV at $4\sigma$ from null hypothesis.
Allowed regions for the characteristic phenomenological parameters 
of the model have been presented.
The obtained allowed 
interval for the mass of the possible mediator of the interaction
is well in agreement with the
typical requirements of the phenomenological analyses available in
literature.

Finally, we further remind that the U boson interpretation is not
the unique one since, for example, there are domains in general SUSY parameter space
where LSP-electron interaction can dominate LSP-quark one.

\vspace{0.6cm}

\appendix

\begin{center}
\LARGE
{\bf APPENDIX} 
\normalsize
\end{center}

\section{$\chi^0$ interaction with atoms}

The inclusive scattering of $\chi^0$ particle 
on an atom $A$ is here analyzed: \mbox{$\chi^0 A \rightarrow \chi^0 X$,}
where $X$ denotes the final state of the atom.
The cross section of the process is obtained by
summing over the possible contributions of all the $X$ final states:
\begin{equation}
d\sigma_{\chi^0 A} \propto
\sum_X \left| T_{AX} \right| ^2 =
\sum_X
\langle A,\chi^0(k) | \chi^0(k'),X \rangle
\langle X,\chi^0(k') | \chi^0(k),A \rangle \; ;
\label{eq:a1}
\end{equation}
here $T_{AX}$ is the transition amplitude when the final state is $X$.

Since it has been assumed that the interaction
of $\chi^0$ with the electrons is dominant,
we can use a full set of electronic plane wavefunctions, $e(p)$, 
and rewrite:
\begin{equation}
\langle A,\chi^0(k) | 
= \sum_p 
\langle A | e(p) \rangle
\langle e(p),\chi^0(k) | 
\label{eq:a2}
\end{equation}
\begin{equation}
|\chi^0(k'),X \rangle 
= \sum_{p'} 
\langle e(p') | X \rangle
| \chi^0(k'),e(p') \rangle \; . 
\label{eq:a3}
\end{equation}
Therefore:
\begin{equation}
T_{AX} =
\sum_{p,p'} 
\langle A | e(p) \rangle
T_{(p+k-p'-k')}
\langle e(p') | X \rangle
\label{eq:a3a}
\end{equation}
where $T_{(p+k-p'-k')} = 
\langle e(p),\chi^0(k) | \chi^0(k'),e(p') \rangle 
\propto M \times \delta(p+k-p'-k') $ is the transition amplitude for
free electron $ - \chi^0$ elastic scattering and
$M$ is the matrix element reported in eq. (\ref{eq:d1}).

Since $X$ is whatever final state:
$\sum_X
\langle e(p') | X \rangle
\langle X | e(p'') \rangle = \delta(p'-p'') $;
therefore, eq. (\ref{eq:a1}) can be written as:
\begin{eqnarray}
\sum_X T^2_{AX} & = &
\sum_{p,p',p'''}
\langle A | e(p) \rangle
T_{(p+k-p'-k')}
T^*_{(p'''+k-p'-k')}
\langle e(p''') | A \rangle \nonumber \\
& \propto &
\sum_{p,p'}
\rho(p) |M|^2 \delta(p+k-p'-k')
\label{eq:a4}
\end{eqnarray}
where $\rho(p) = | \langle A | e(p) \rangle |^2$ is the 
momentum distribution function of the electrons in the atom $A$. 
Finally, we can deduce $d\sigma_{\chi^0 A}= d\sigma_{\chi^0 e} \cdot \rho(p) d^3 p$,
where $d\sigma_{\chi^0 e}$ is the $\chi^0-e^-$ elastic scattering cross section
given in eq. (\ref{eq:d1}).

\section{The invariant amplitude for $\chi^0-e^-$ elastic scattering}

In the following we consider the elastic scattering 
of the $\chi^0$ fermion on electron by using a Fermi-like 
4-fermion contact interaction.

\subsection{The VA subcase}

The squared matrix element, averaged over the initial spins and summed over the final ones, 
can be written as:

\begin{equation}
\overline{|M_{VA}|^2} = 
G^2 L^{\mu\nu}_{(\chi^0)}L_{\mu\nu}^{(e)},
\label{eq:VA1}
\end{equation}
where:
\begin{equation}
L^{\mu\nu}_{(\chi^0)}=\frac{1}{2} \sum_{spin}
\left[ \bar{U}_{\chi^0}(k') \gamma^{\mu} (g_V+g_A \gamma^5)
U_{\chi^0}(k) \right] 
\left[ \bar{U}_{\chi^0}(k) \gamma^{\nu} (g_V+g_A \gamma^5)
U_{\chi^0}(k') \right]
\label{eq:VA2}
\end{equation}
\begin{equation}
L_{\mu\nu}^{(e)}=\frac{1}{2} \sum_{spin}
\left[ \bar{U}_{e}(p') \gamma_{\mu} (c_V+c_A \gamma^5)
U_{e}(p)\right] 
\left[ \bar{U}_{e}(p) \gamma_{\nu} (c_V+c_A \gamma^5)
U_{e}(p') \right] \; .
\label{eq:VA3}
\end{equation}

Let us focus just on eq. (\ref{eq:VA2}), since eq. (\ref{eq:VA3})
has the same structure. One can write:
\begin{eqnarray}
L^{\mu\nu}_{(\chi^0)}&=&\frac{1}{2} Tr
\left[ (\not \!k'+m_{\chi^0}) \gamma^{\mu} (g_V+g_A \gamma^5)
 (\not \!k + m_{\chi^0}) \gamma^{\nu} (g_V+g_A \gamma^5) \right] \nonumber \\
&=&T^{AA}+T^{VA}+T^{AV}+T^{VV} 
\label{eq:VA4}
\end{eqnarray}
The four terms can be explicited as:
\begin{equation}
T^{AA}=\frac{1}{2} Tr
\left[ (\not \!k'+m_{\chi^0}) \gamma^{\mu} g_A \gamma^5
 (\not \!k + m_{\chi^0}) \gamma^{\nu} g_A \gamma^5 \right]
\label{eq:VA5}
\end{equation}
\begin{equation}
T^{VV}=\frac{1}{2} Tr
\left[ (\not \!k'+m_{\chi^0}) \gamma^{\mu} g_V 
 (\not \!k + m_{\chi^0}) \gamma^{\nu} g_V \right]
\label{eq:VA6}
\end{equation}
\begin{equation}
T^{AV}=\frac{1}{2} Tr
\left[ (\not \!k'+m_{\chi^0}) \gamma^{\mu} g_A \gamma^5
 (\not \!k + m_{\chi^0}) \gamma^{\nu} g_V \right]
\label{eq:VA7}
\end{equation}
\begin{equation}
T^{VA}=\frac{1}{2} Tr
\left[ (\not \!k'+m_{\chi^0}) \gamma^{\mu} g_V
 (\not \!k + m_{\chi^0}) \gamma^{\nu} g_A \gamma^5 \right]
\label{eq:VA8}
\end{equation}

\noindent By using trace theorems one gets:
\begin{equation}
T^{AA}=\frac{1}{2} g_A^2 Tr
\left[ \not \!k'\gamma^{\mu}\not \!k \gamma^{\nu}
- m^2_{\chi^0} \gamma^{\mu} \gamma^{\nu}  \right]=2 
g_A^2(k'^{\mu}k^{\nu}+k'^{\nu}k^{\mu}-k'kg^{\mu\nu}- m_{\chi^0}^2 
g^{\mu\nu})
\label{eq:VA9}
\end{equation}
\begin{equation}
T^{VV}=\frac{1}{2}g_V^2 Tr
\left[ \not \!k'  \gamma^{\mu} \not \!k  \gamma^{\nu}
+ m_{\chi^0}^2 \gamma^{\mu}  \gamma^{\nu}\right]
= 2 g_V^2 (k'^{\mu}k^{\nu}+k'^{\nu}k^{\mu}-k'kg^{\mu\nu} + m_{\chi^0}^2
g^{\mu\nu})
\label{eq:VA10}
\end{equation}
\begin{equation}
T^{AV}=\frac{1}{2} Tr
\left[ (\not \!k'+m_{\chi^0}) \gamma^{\mu}
 (\not \!k - m_{\chi^0}) \gamma^{\nu}g_A \gamma^5 g_V \right]
\label{eq:VA11}
\end{equation}
\begin{equation}
T^{VA}=\frac{1}{2} g_V  g_A Tr
\left[ \gamma^5 \not \!k' \gamma^{\mu} 
\not \!k \gamma^{\nu} \right]=T^{AV}
\label{eq:VA12}
\end{equation}
\begin{equation}
T^{VA}+T^{AV}=-g_V  g_A 
4i\varepsilon^{\alpha \mu \beta \nu }
k'_{\alpha} k_{\beta}
\label{eq:VA13}
\end{equation}

Thus, one can write:
\begin{eqnarray}
L^{\mu\nu}_{(\chi^0)} &=&
2(g_V^2+g_A^2)\left[
k'^{\mu}k^{\nu}+k'^{\nu}k^{\mu}-k'kg^{\mu\nu} \right] + \nonumber \\
&& +2(g_V^2-g_A^2)m_{\chi^0}^2 g^{\mu\nu}
-4 g_V  g_A
i\varepsilon^{\alpha \mu \beta \nu }
k'_{\alpha} k_{\beta} 
\label{eq:VA14}
\end{eqnarray}

Finally, the matrix element for the process can be written as:
\begin{eqnarray}
\overline{|M_{VA}|^2} = 8G^2 \left[ A (p'k')(pk) + 
B(p'k)(pk')-C(kk')m_e^2
-D (pp') m_{\chi^0}^2
\right] ,
\label{eq:VA19}
\end{eqnarray}
where:
\begin{eqnarray}
A &=& (g_V^2+g_A^2)(c_V^2+c_A^2) +4g_Vg_Ac_Vc_A = (c_Vg_V+c_Ag_A)^2+(c_Vg_A+c_Ag_V)^2 \nonumber \\
B &=& (g_V^2+g_A^2)(c_V^2+c_A^2) -4g_Vg_Ac_Vc_A = (c_Vg_V-c_Ag_A)^2+(c_Vg_A-c_Ag_V)^2 \nonumber \\
C &=& (g_V^2+g_A^2)(c_V^2-c_A^2) \nonumber \\
D &=& (g_V^2-g_A^2)(c_V^2+c_A^2)
\label{eq:VA20}
\end{eqnarray}

In the case of $V \pm A$ interaction ($|c_V|=|c_A|$ and $|g_V|=|g_A|$)
the matrix element is:
\begin{eqnarray}
\overline{|M_{V \pm A}|^2} = 8G^2 \left[ A (p'k')(pk) + 
B(p'k)(pk') \right]
\label{eq:V-A}
\end{eqnarray}
\noindent knowing that $\chi^0$ is not relativistic
(see text), one obtains:
$(p'k')(pk) \simeq p'_0k'_0p_0k_0$
and 
$(p'k)(pk') \simeq p'_0k'_0p_0k_0$;
moreover for $E_d \sim keV$ one has $p'_0 \simeq p_0$, giving:
\begin{eqnarray}
\overline{|M_{V \pm A}|^2} \simeq 16G_{V \pm A}^2 m_{\chi^0}^2p_0^2 ,
\label{eq:V-A2}
\end{eqnarray}
\noindent where the Fermi effective coupling constant is:
$G_{V \pm A}^2 = 
G^2(c_V^2+c_A^2)(g_V^2+g_A^2)$.
For this particular case, the dependence on
$v_{\chi^0}$ can be neglected, while the dependence
on $p$ are included in: $p_0^2 = p^2+m_e^2$.

\subsection{The SP subcase}

Similarly as above, one has:
\begin{equation}
\overline{|M_{SP}|^2} =  
G^2 L_{(\chi^0)}L_{(e)}
\label{eq:SP1}
\end{equation}
\begin{equation}
L_{(\chi^0)}=\frac{1}{2} \sum_{spin}
\left[ \bar{U}_{\chi^0}(k') (g_S+ig_P \gamma^5)
U_{\chi^0}(k) \right] 
\left[ \bar{U}_{\chi^0}(k) (g_S+ig_P \gamma^5)
U_{\chi^0}(k') \right] 
\label{eq:SP2}
\end{equation}
\begin{eqnarray}
L_{(\chi^0)}&=&\frac{1}{2} Tr
\left[ (\not \!k'+m_{\chi^0}) (g_S+ig_P \gamma^5)
(\not \!k + m_{\chi^0}) (g_S+ig_P \gamma^5) \right] \nonumber \\
&=&T^{SS}+T^{SP}+T^{PS}+T^{PP} 
\label{eq:SP3}
\end{eqnarray}
There:
\begin{equation}
T^{SS}=\frac{1}{2} g_S^2 Tr
\left[ (\not \!k'+m_{\chi^0}) (\not \!k + m_{\chi^0}) 
\right]=2g_S^2(k'k+m_{\chi^0}^2) 
\label{eq:SP4}
\end{equation}
\begin{equation}
T^{PP}= - \frac{1}{2} g_P^2 Tr
\left[ (\not \!k'+m_{\chi^0})\gamma^5 (\not \!k + m_{\chi^0})\gamma^5 
\right]=2g_p^2(k'k-m_{\chi^0}^2)
\label{eq:SP5}
\end{equation}
\begin{equation}
T^{PS}=  \frac{1}{2} ig_Pg_S Tr
\left[ (\not \!k'+m_{\chi^0})\gamma^5 (\not \!k + m_{\chi^0}) 
\right]
\label{eq:SP6}
\end{equation}
\begin{equation}
T^{SP}+T^{PS}= ig_Pg_S Tr
\left[ (\not \!k'+m_{\chi^0})\gamma^5 m_{\chi^0} \right]
=0
\label{eq:SP7}
\end{equation}
Hence:
\begin{equation}
L_{(\chi^0)}=2\left[(g_S^2+g_P^2) k'k + (g_S^2-g_P^2)m_{\chi^0}^2 \right]=
2(g_+k'k+g_-m_{\chi^0}^2) ,
\label{eq:SP8}
\end{equation}
where $g_+=g_S^2+g_P^2>0$ and $g_-=g_S^2-g_P^2$.
\vspace{0.3cm}
Finally:
\begin{equation}
\overline{|M_{SP}|^2} =  
4G^2 \left[g_+c_+(k'k)(p'p)+g_+c_-(k'k)m_e^2
+g_-c_+(p'p)m_{\chi^0}^2+g_-c_-m_{\chi^0}^2m_e^2
\right]
\label{eq:SP9}
\end{equation}

In the particular pure scalar case
($g_P = c_P = 0$) one obtains:
\begin{equation}
\overline{|M_{S}|^2} =  
4G^2 g_S^2 c_S^2 \left[(k'k)+m_{\chi^0}^2\right]\left[(p'p)+m_e^2
\right]\simeq 8 G^2_S m_{\chi^0}^2 \left[p'_0p_0-\vec{p'}\vec{p}+m_e^2
\right]
\label{eq:SP92}
\end{equation}
Thus, considering the momentum distribution of atomic electron, for
$E_d \sim keV$ practically $\vec{p'} \sim - \vec{p}$ and, therefore: 
\begin{equation}
\overline{|M_{S}|^2} \sim  
16 G^2_S m_{\chi^0}^2 p_0^2
\label{eq:SP93}
\end{equation}

\noindent where the Fermi effective coupling constant is:
$G_{S}^2 = 
G^2 c_S^2 g_S^2$.

Also in this case there is a negligible dependence 
from $v_{\chi^0}$ and a weak dependence from $p$.


\begin{thebibliography}{99}
\itemsep -2pt

\bibitem{Uboson}   Y. Ascasibar, P. Jean, C. Boehm and J. Knoedlseder,
                   Mon. Not. Roy. Astron. Soc. 368 (2006) 1695;
                   C. Jacoby and S. Nussinov, JHEP 05 (2007) 017.
\bibitem{Uboson2}  D.P. Finkbeiner and N. Weiner, Phys. Rev. D 76 (2007) 083519.
\bibitem{Uboson3}  M. Pospelov and A. Ritz, Phys. Lett. B 651 (2007) 208, hep-ph/0703128.
\bibitem{fay}      P. Fayet, Phys. Rev. D 75 (2007) 115017.
\bibitem{integral} J. Knodlseder et al., Astron. Astrophys. 513 (2005) 441;
                   P. Jean et al., Astron. Astrophys. L55 (2003) 407; 
                   J. Knodlseder et al., Astron. Astrophys. L457 (2003) 411.
\bibitem{theor}    C. Boehm and Y. Ascasibar, Phys. Rev. D 70 (2004) 115013;
                   G. Weidenspointner et al., astro-ph/0702621.
\bibitem{B200}     B. Tatischeff and E. Tomasi-Gustafsson, arXiv:0710.1796 and arXiv:0710.1798. 
\bibitem{B200a}    N.G. Deshpande, G.Eilam and J. Jiang, Phys. Lett. B 632 (2006) 212;
                   D.S. Gorbunov and V.A. Rubakov, Phys. Rev. D 73 (2006) 035002;
                   C.H. Chen et al., arXiv:0708.0937.
\bibitem{B20}      T. Walcher, hep-ph/0111279.
\bibitem{B10}      F.W.N. de Boer et al., J. Phys. G 27 (2001) L29; J. Phys. G 23 (1997) L85;
                   Nucl. Phys. B 72 (1999) 189; M. El-Nadi and O.E. Badawy, 
                   Phys. Rev. Lett. 61 (1988) 1271;
                   K. Asakimori et al., J. Phys. G 25 (1999) L133.
\bibitem{patisalam}       R. Foot, H. Lew and R.R. Volkas, Phys. Rev. D 44 (1991) 859.
\bibitem{quartification}  K.S. Babu, E. Ma and S. Willenbrock, Phys. Rev. D 69 
                          (2004) 051301(R); 
                          S.L. Chen and E. Ma, Mod. Phys. Lett. A 19 (2004) 1267;
                          A. Demaria, C.I. Low, R.R. Volkas, Phys. Rev. D 72 (2005) 075007.
\bibitem{lcolor}   R. Foot and H. Lew, Phys. Rev. D 41 (1990) 3502;
                   R. Foot, H. Lew and R.R. Volkas, Phys. Rev. D 44 (1991) 1531;
                   R. Foot and R.R. Volkas, Phys. Lett. B 645 (2007) 345;
                   K.S. Babu, T.W. Kephart, H. Pas, arXiv:0709.0765 [hep-ph].
\bibitem{RNC}      R. Bernabei el al., La Rivista del Nuovo Cimento 26 n.1 (2003) 1-73.
\bibitem{ijmd}     R. Bernabei el al., Int. J. Mod. Phys. D 13 (2004) 2127.
\bibitem{epj06}    R. Bernabei et al., Eur. Phys. J. C. 47 (2006) 263.
\bibitem{ijma2}    R. Bernabei el al., Int. J. Mod. Phys. A 22 (2007) 3155-3168. 
\bibitem{ijma}     R. Bernabei et al., Int. J. Mod. Phys. A 21 (2006) 1445.
\bibitem{chann}    R. Bernabei et al., to appear on Eur. Phys. J. C, arXiv:0710.0288.
\bibitem{Bo03}     A. Bottino et al., Phys. Rev. D 67 (2003) 063519;
                   A. Bottino et al., Phys. Rev. D 68 (2003) 043506.
\bibitem{Bo04}     A. Bottino et al., Phys. Rev. D 69 (2004) 037302.
\bibitem{Botdm}    A. Bottino et al., Phys. Lett. B 402 (1997) 113;
                                      Phys. Lett. B 423 (1998) 109;
                                      Phys. Rev. D 59 (1999) 095004; 
                                      Phys. Rev. D 59 (1999) 095003;
                                      Astrop. Phys. 10 (1999) 203;
                                      Astrop. Phys. 13 (2000) 215;
                                      Phys. Rev. D 62 (2000) 056006;
                                      Phys. Rev. D 63 (2001) 125003;
                                      Nucl. Phys. B 608 (2001) 461.
\bibitem{khlopov}  K. Belotsky, D. Fargion, M. Khlopov and R.V. Konoplich, hep-ph/0411093.
\bibitem{Wei01}    D. Smith and N. Weiner, Phys. Rev. D 64 (2001) 043502;
                   D. Tucker-Smith and N. Weiner, Phys. Rev. D 72 (2005) 063509.
\bibitem{foot}     R. Foot, hep-ph/0308254.
\bibitem{Saib}     S. Mitra, Phys. Rev. D 71 (2005) 121302(R).
\bibitem{droby1}   E.M. Drobyshevski et al., arXiv:0704.0982
\bibitem{droby2}   E.M. Drobyshevski, arXiv:0706.3095
\bibitem{sneu}     C. Arina and N. Fornengo, arXiv:0709.4477
\bibitem{zoom}     A. Bottino et al., arXiv:0710.0553
\bibitem{Hep}      P. Belli et al., Phys. Rev. D 66 (2002) 043503.
\bibitem{biggs}    F. Biggs et al., Atomic data and nuclear data tables 16 (1975) 201.
\bibitem{cprof2}   D. Brusa et al., Nucl. Inst. \& Meth. A  379 (1996) 167.
\bibitem{cprof3}   R. Ribberfors et al., Phys. Rev. A 26 (1982) 3325;
                                         Phys. Rev. B 12 (1975) 2067.

\end{thebibliography}
\end{document}